\documentclass[twocolumn,aps,nolongbibliography]{revtex4-2}

\usepackage{graphicx}
\usepackage{amssymb}
\usepackage{amsfonts}
\usepackage{amsmath}
\usepackage{amsbsy}
\usepackage{natbib}
\usepackage{comment}

\usepackage{color}
\usepackage{float}

\usepackage[colorlinks=true, urlcolor=blue, citecolor=blue, linkcolor=blue]{hyperref}

\renewcommand{\vec}[1]{\mbox{\boldmath$\mathrm{#1}$}}
\let\sb=_ \catcode`\_=\active \def_#1{\ensuremath \sb{\rm#1}}

\renewcommand{\vec}[1]{\mbox{\boldmath$\mathrm{#1}$}}
\newcommand{\be}{\begin{equation}}
\newcommand{\ee}{\end{equation}}
\newcommand{\ben}{\begin{eqnarray}}
\newcommand{\een}{\end{eqnarray}}

\begin{document}


\title{Steering skyrmions with microwave and THz  electric pulses}

\author{Xi-guang Wang$^1$, Guang-hua Guo$^1$, V. K. Dugaev$^{2}$, J. Barna\'{s}$^{3,4}$, J. Berakdar$^5$, S. S. P. Parkin$^{6}$, A. Ernst$^{6,7}$, L. Chotorlishvili$^{2}$}
\address{$^1$ School of Physics and Electronics, Central South University, Changsha 410083, China\\
$^2$ Department of Physics and Medical Engineering, Rzeszow University of Technology, 35-959 Rzeszow, Poland\\
$^3$ Faculty of Physics, Adam Mickiewicz University, ul. Uniwersytetu Poznanskiego 2, 61-614 Poznan, Poland\\
$^4$ Institute of Molecular Physics, Polish Academy of Sciences, ul. M. Smoluchowskiego 17, 60-179 Pozna\'{n}, Poland\\
$^5$ Institut f\"ur Physik, Martin-Luther Universit\"at Halle-Wittenberg, D-06120 Halle/Saale, Germany \\
$^6$ Max Planck Institute of Microstructure Physics, Weinberg 2, D-06120 Halle, Germany\\
$^{7}$ Institute for Theoretical Physics, Johannes Kepler University, Altenberger Stra\ss e 69, 4040 Linz, Austria}

\date{\today}

\begin{abstract}
Tools for controlling electrically the motion of magnetic skyrmions  are important elements towards their use in spintronic devices. Here, we propose and demonstrate
the transport of   skyrmions via GHz and THz  electric pulses. The method relies on using  polarization textured  pulses such that the skyrmion experiences  (via its inherent magnetoelectricity)
 the out-of-plane and in-plane components of the pulse electric field. It is shown how the electric  field drags efficiently  the skyrmion. The   control of  the skyrmion motion depends solely  on the amplitude of
electric fields, frequency, polarization or  phase in case two pulses are applied. Micromagnetic calculations  supported by analytic modeling and analysis indicate the experimental feasibility of the  control scheme.

\end{abstract}

\maketitle

\section{Introduction}

Magnetic skyrmions are topological textures formed in thin magnetic films with no inversion symmetry. Their unique physical features, such as nanoscale size, high stability, and high mobility,
make them ideal    for diverse applications in future spintronic devices \cite{muhlbauer2009skyrmion, Yu2010,doi:10.1063/5.0042917, Tomasello2014,Zzvorka2019, Zhang2015,Fert2013, Wiesendanger2016, PhysRevLett.129.126101, PhysRevB.106.104424}. The high mobility of skyrmions reduces the energy costs of skyrmion-based transfer of information and manipulation of skyrmion-based memory elements. Recent studies  showed that skyrmion displacement can be effectively driven by electric/magnonic spin transfer torques \cite{PhysRevLett.107.136804,PhysRevLett.110.207202,Iwasaki2013,PhysRevB.89.064412,PhysRevB.91.104435,Zhang2016,Tomasello2014,Goebel2019,PhysRevLett.120.237203}, magnetic field gradient \cite{Wang2017,Zhang2018,PhysRevB.86.054432}, voltage \cite{PhysRevApplied.11.014004}, microwave magnetic field \cite{Moon2016,PhysRevB.92.020403}, or thermal gradient \cite{PhysRevB.86.054432,PhysRevLett.112.187203,PhysRevLett.111.067203}.   Recently it has been reported that an electric field gradient can be used for manipulating skyrmions \cite{PhysRevB.87.100402, PhysRevB.92.134411, PhysRevB.99.064426}. The inherent spin non-collinearity renders skyrmions magnetoelectric.
Exploiting magnetoelectric coupling enables an  optical tweezing  of skyrmions and vortices \cite{Wang2020, PhysRevLett.125.227201,PhysRevApplied.16.034032}, even though near  intense focused  fields are necessary.

In this paper  we propose a new protocol to move skyrmions with  propagating electric field pulses. The fields are polarized such that they possess a component in the plane  and a component orthogonal to the plane of the   skyrmion. The speed and  direction of the skyrmion motion are determined by the amplitudes, frequencies, and phases of the  electric pulses. This mechanism is different from the skyrmion motion driven by the microwave magnetic field, where an asymmetric skyrmion distortion induced by an in-plane static magnetic field is necessary \cite{PhysRevB.92.020403}. Our mechanism is operational when using harmonic as well as for two broad-band THz time-asymmetric  pulses with perpendicular components. The  pulses are applied uniformly to the sample, and field focusing at the nanoscale is not necessary. Our results point to new opportunities for the optical control of skyrmions.

\section{Theoretical model}

\begin{figure}
	\includegraphics[width=0.5\textwidth]{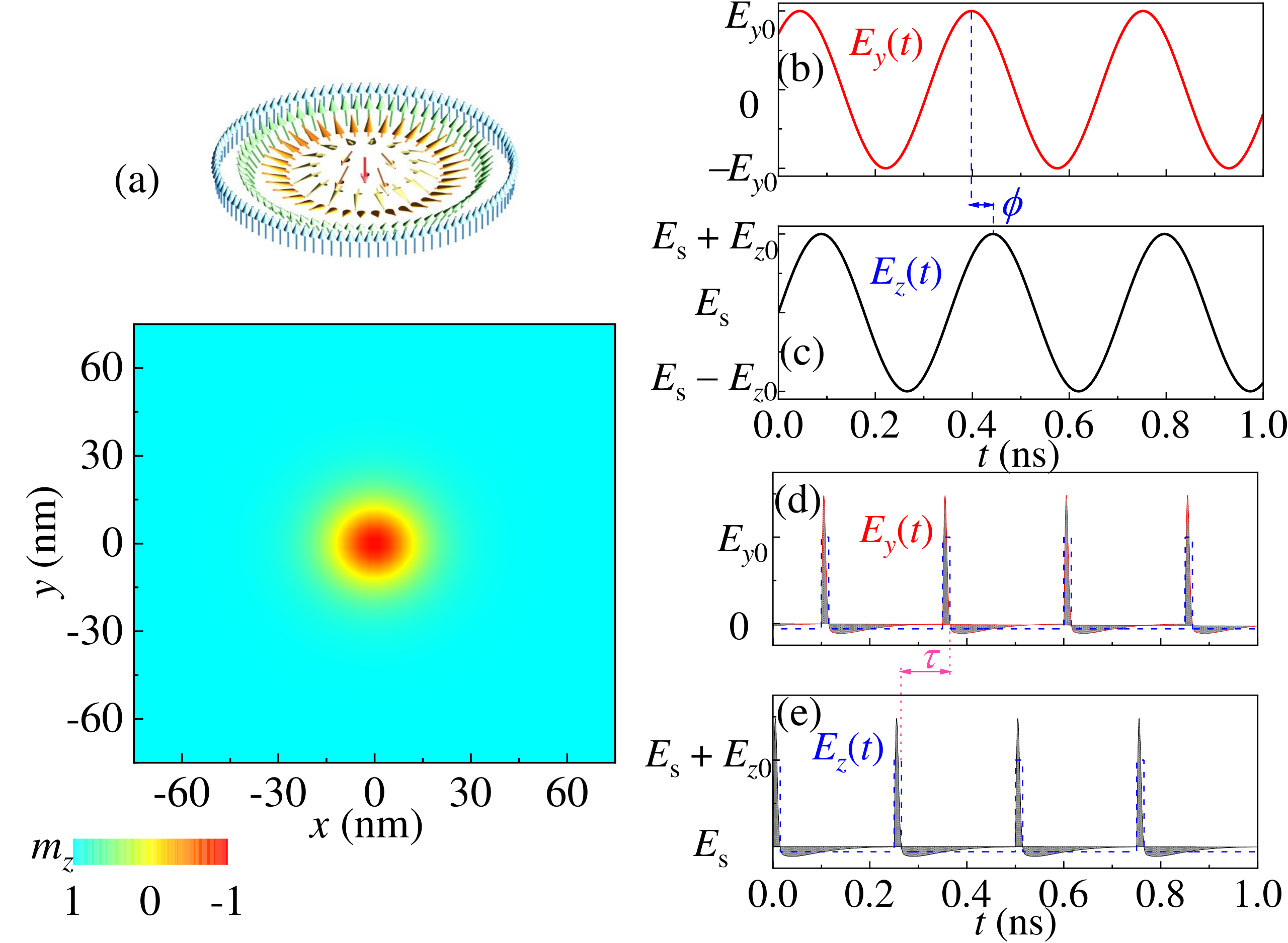}
	\caption{\label{model} (a) Magnetization configuration of skyrmion structure and its $ z $ component $ m_z $ profile. (b-c) $ E_y $ and $ E_z $ components of GHz microwave electric fields. (d-e) Two periodic time-asymmetric pulses $ E_y(t) $ and $ E_z(t) $. The red (d) and black (e) envelopes show the time profile of an experimentally feasible asymmetric pulse, and the blue dotted line describes the shape of the field adopted in numerical simulations.}
\end{figure}

To study the skyrmion motion governed by an electric field, we consider  a  magnetic system with the free energy containing the exchange, Zeeman, and magnetoelectric (ME) interaction terms
\begin{equation}
\displaystyle F[\vec{m}] = \int[A_{ex}(\vec{\nabla m})^2 - \mu_0 M_s m_z H_z + E_{me}]d\vec{r}.
\label{energy}
\end{equation}
Here, $ \vec{M} = M_s \vec{m} $, $ M_s $ is the saturation magnetization, $ A_{ex} $ is the exchange constant, and $ H_z $ is the external magnetic field applied along $ \textbf{z} $ direction. The ME interaction  allows a coupling to an electric field $ \vec{E} $ via  the effective electric polarization $ \vec{P} = c_E [(\vec{m} \cdot \vec{\nabla})\vec{m} - \vec{m}(\vec{\nabla} \cdot \vec{m})] $, which is associated with the nonuniform magnetic distribution within the skyrmion.  The ME coupling parameter $ c_E $ plays the role of an effective Dzyaloshinskii-Moriya (DM) constant $ D = c_E |\vec{E}| $.\cite{PhysRevLett.106.247203, PhysRevLett.95.057205} The skyrmion dynamics is governed by the Landau-Liffhitz-Gilbert (LLG) equation,
 \begin{equation}
 \displaystyle \frac{\partial \vec{m}}{\partial t} = - \gamma \vec{m} \times \vec{H}_{\mathrm{\rm eff}}+ \alpha \vec{m} \times \frac{\partial \vec{m}}{\partial t},
 \label{LLG}
 \end{equation}
 where $ \gamma $ is the gyromagnetic ratio, and $ \alpha $ is the phenomenological Gilbert damping constant. The effective field $ \vec{H}_{\mathrm{\rm eff}} = -\frac{1}{\mu_0 M_s} \frac{\delta F}{\delta \vec{m}}$ consists of the exchange field, the applied external magnetic field, and the effective DM field. The total electric field $ \vec{E} = \vec{E}_s + \vec{E}(t) $ includes a perpendicular static electric field $ \vec{E}_s $ and a time-varying electric field $ \vec{E}(t) $. Due to the ME term $c_E$,  $ \vec{E}_s =  (0, 0, E_s) $ acting the electric polarization $\textbf{P}$  changes the energy by $-\textbf{E}_s \cdot \textbf{P}$. This coupling  can be viewed as an effective DM interaction with the constant $ D_0 = c_E E_s $. The effective DM term  stabilizes a skyrmion magnetic structure  of N$ {\rm \acute{e}} $el type (Fig. \ref{model}(a)). The impact of the static field $ \vec{E}_s $ is similar to a breaking of inversion symmetry of the lattice structures such as B20 and ferromagnet/heavy metal bilayers \cite{PhysRevLett.124.057201,muhlbauer2009skyrmion,Yu2010, Soumyanarayanan2017}.
 The time-dependent field $ \vec{E}(t) $ is applied to the whole sample to trigger the skyrmion dynamics. Here, we consider two types of  $ \vec{E}(t) $: (i) a harmonic pulse $ \vec{E}(t) = E_{z0} \sin(2 \pi f t) \vec{e}_z +  E_{y0} \sin(2 \pi f t + \phi) \vec{e}_y  $ with a frequency $ f $ (Fig. \ref{model}(b-c)), and (ii) periodic time-asymmetric broad-band  THz pulses (i.e., half-cycle laser pulses, see Fig. \ref{model}(d-e)) with the period $ T $ and delay time $ \tau $. In the following analysis, the frequencies $ f $ and $ 1/T $ are in the GHz range, and the length of the laser pulse head is 15 ps. Such THz pulses were used to control the vortex dynamics even in metallic samples \cite{Yu2020} but here we consider insulting samples.

In numerical calculations we assume the following parameters: the saturation magnetization $ M_s = 1.4 \times 10^5 $ A/m, the exchange constant $ A_{ex} = 3 \times 10^{-12} $ J/m, the ME coupling strength $ c_E = 0.9 $ pC/m, and the Gilbert damping constant $ \alpha = 0.01 $. The N$ {\rm \acute{e}} $el skyrmion is stabilized by the constant electric field $ E_s = 2.5 $ MV/cm and $ H_z = 1 \times 10^5 $ A/m. The finite difference simulations based on Eq. (\ref{LLG}) are done for the magnetic film with the size of $ 600{\rm nm} \times 600{\rm nm} \times 10{\rm nm} $, and the $ 3{\rm nm} \times 3{\rm nm} \times 10{\rm nm} $ cell size is adopted. From the simulated profile $ \vec{m}(x,y) $, we extract the topological charge density $ c(x,y) = (1/4\pi)\vec{m}\cdot(\partial_x \vec{m} \times \partial_y \vec{m}) $, and the total topological charge of a single skyrmion is $ Q = \int c\, dx dy = -1 $. In the following simulations, the single skyrmion starts to move under the effect of time-varying electric field, and by defining the skyrmion position $ \vec{q} = (q_x, q_y) $ weighed by the topological charge $ \vec{q} = \int d^2\vec{r}\, (c \vec{r}) / Q  $, we  characterize the skyrmion motion by $ \vec{q}(t) $.

\begin{figure}
	\includegraphics[width=0.50\textwidth]{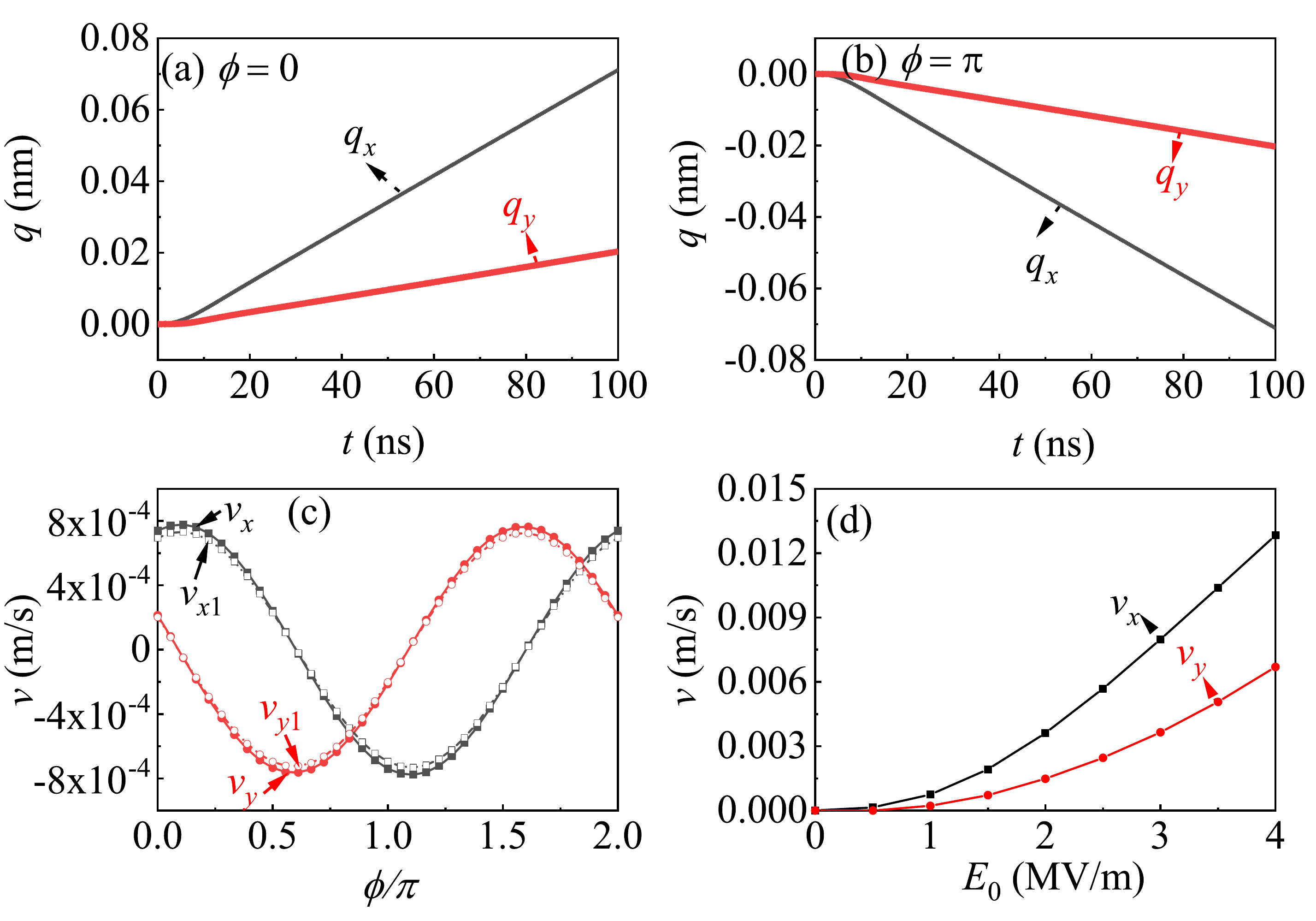}
	\caption{\label{phasevxy} (a-b) Center of the skyrmion $ (q_x, q_y) $ steered through the applied microwave electric fields ($ E_{z0} = E_{y0} = E_0 = 1 $ MV/m, $ f = 2.825 $ GHz, (a) $  \phi = 0 $ and (b) $  \phi = \pi $). (c-d) The velocity $ (v_x, v_y) $ as a function of the phase difference $ \phi $ (c) and amplitude $ E_0 $ (d). The solid dot lines are obtained from micromagnetic simulations, while the open dot lines are estimated from Eq. (\ref{velocity}).}
\end{figure}

\section{Skyrmion motion driven by microwave electric field}

Considering the skyrmion motion driven by GHz  electric fields with two $ y $ and $ z $ components,
when only one  component $ y $ (or $ z $) is applied, the motion of the skyrmion is not continuous. Figure \ref{phasevxy} shows the displacement of a single skyrmion for $ E_{z0} = E_{y0} = E_0 = 1 $ MV/m, $ f = 2.825 $ GHz, and $  \phi = 0 $. As  evident, both $ x $ and $ y $ components of $ \vec{q} = (q_x, q_y)$ change  in time, and the speed $ v_x = 7.4 \times 10^{-4} $ m/s along $ x $ is higher than $ v_y = 2.1 \times 10^{-4} $ m/s along $ y $. The velocity of the skyrmion depends on the phase difference $ \phi $. In Fig. \ref{phasevxy}(b), the directions of $ v_x $ and $ v_y $ are reversed by changing the phase to $  \phi = \pi $. The relation between the skyrmion velocity $ \vec{v} $ and the phase difference $ \phi $ is shown in Fig. \ref{phasevxy}(c).  Apparently, the magnitude of $ |\vec{v}| = \sqrt{v_x^2 + v_y^2} $ is a constant and does not depend on $ \phi $.  One component of the velocity approaches its maximum when the second component tends to zero. The velocities can be enhanced by increasing the field amplitude  $ E_{0}$. As follows from  Fig. \ref{phasevxy}(d),  $ |\vec{v}|$ increases quadratically with the microwave field amplitude, indicating a linear dependence on the combination of $ E_{y0} \times E_{z0}$. We conclude that a GHz microwave electric field with a certain frequency moves the skyrmion effectively. The direction of the skyrmion motion can be controlled through the differences in the phase between $ y $ and $ z $ components, and the microwave field amplitude enhances the speed.

\begin{figure}
	\includegraphics[width=0.50\textwidth]{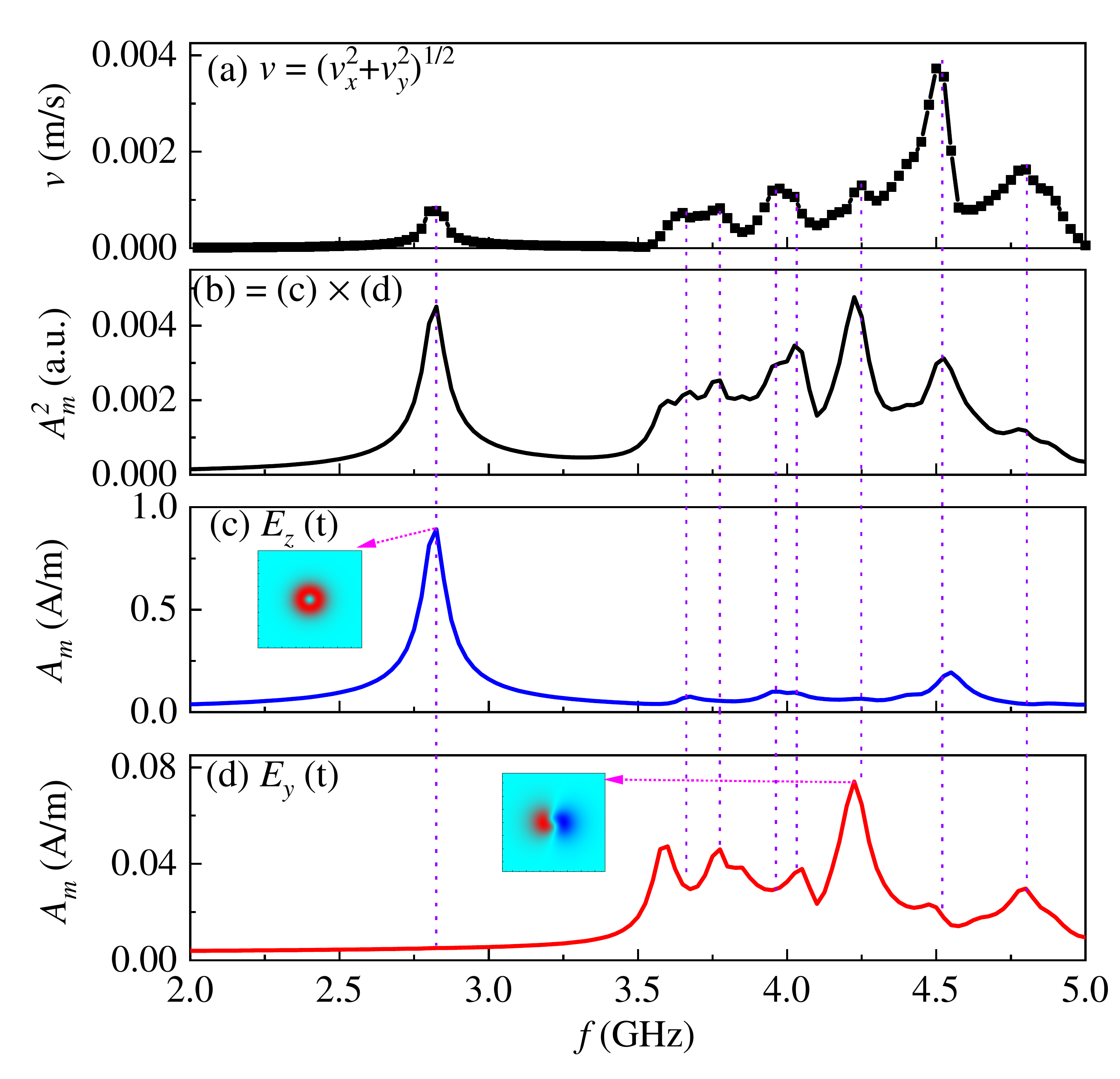}
	\caption{\label{fre} (a) The skyrmion speed $ v = (v_x^2 + v_y^2)^{\frac{1}{2}} $ as a function of the   electric field frequency $ f $. (b-d) The frequency spectra obtained from the Fourier-analysis of the magnetization component $ m_z $. Oscillations are excited by applying a periodic electric field (c) $ E_z(t) = E_0 {\rm sin}(2 \pi f_c t) $ and (d) $ E_y(t) = E_0 {\rm sin}(2 \pi f_c t)$ with $ E_0 = 5 $ MV/m and $ f_c = 20 $ GHz. The curve in (b) is the product of the results shown in (c) and (d). The insets show the spatial distributions of the excited oscillation amplitudes. }
\end{figure}

By steering the frequency $ f $ of microwave electric field $ \vec{E}(t) $ (Fig. \ref{fre}), we find that the skyrmion speed $ |\vec{v}| $ is frequency dependent and several peaks appear on the $ |\vec{v}| $  versus $ f $ curve. These peaks originate from the excitation of internal modes of the skyrmion texture. To study the excitation mode of the skyrmion, we calculate the oscillation spectrum. Applying a periodic pulse $ E_{z(y)}(t) = E_{z0(y0)} \sin(\omega_s t) / (\omega_s t)$ along $ z $ (or $ y $) axis, we compute the Fourier transform of the magnetization oscillations, depicted  in Fig. \ref{fre}(c-d).  The oscillation excited by $ E_z(t) $  (inset of Fig. \ref{fre}(c)) is located in the skyrmion outer boundary and is synchronized in  phase with the external field, i. e., the typical feature of a breathing mode \cite{PhysRevB.87.100402, PhysRevLett.108.017601, PhysRevB.97.064403,miao101063}. The lowest-order breathing mode with 2.825 GHz frequency cannot propagate outside the skyrmion texture. At higher frequencies, 3.68 GHz or 4.55 GHz, the propagating higher order breathing modes are reflected from the geometric boundary and generate standing wave modes. As for the mode excited by $ E_y(t) $, the oscillation in the boundary of a skyrmion is divided into two parts with opposite phases, corresponding to an asymmetric mode (inset of Fig. \ref{fre}(d)). Propagating asymmetric modes generate standing waves in the finite magnetic film, leading to higher-order peaks in the spectrum \cite{PhysRevB.87.100402, PhysRevLett.108.017601, PhysRevB.97.064403,miao101063}.

\begin{figure}
	\includegraphics[width=0.50\textwidth]{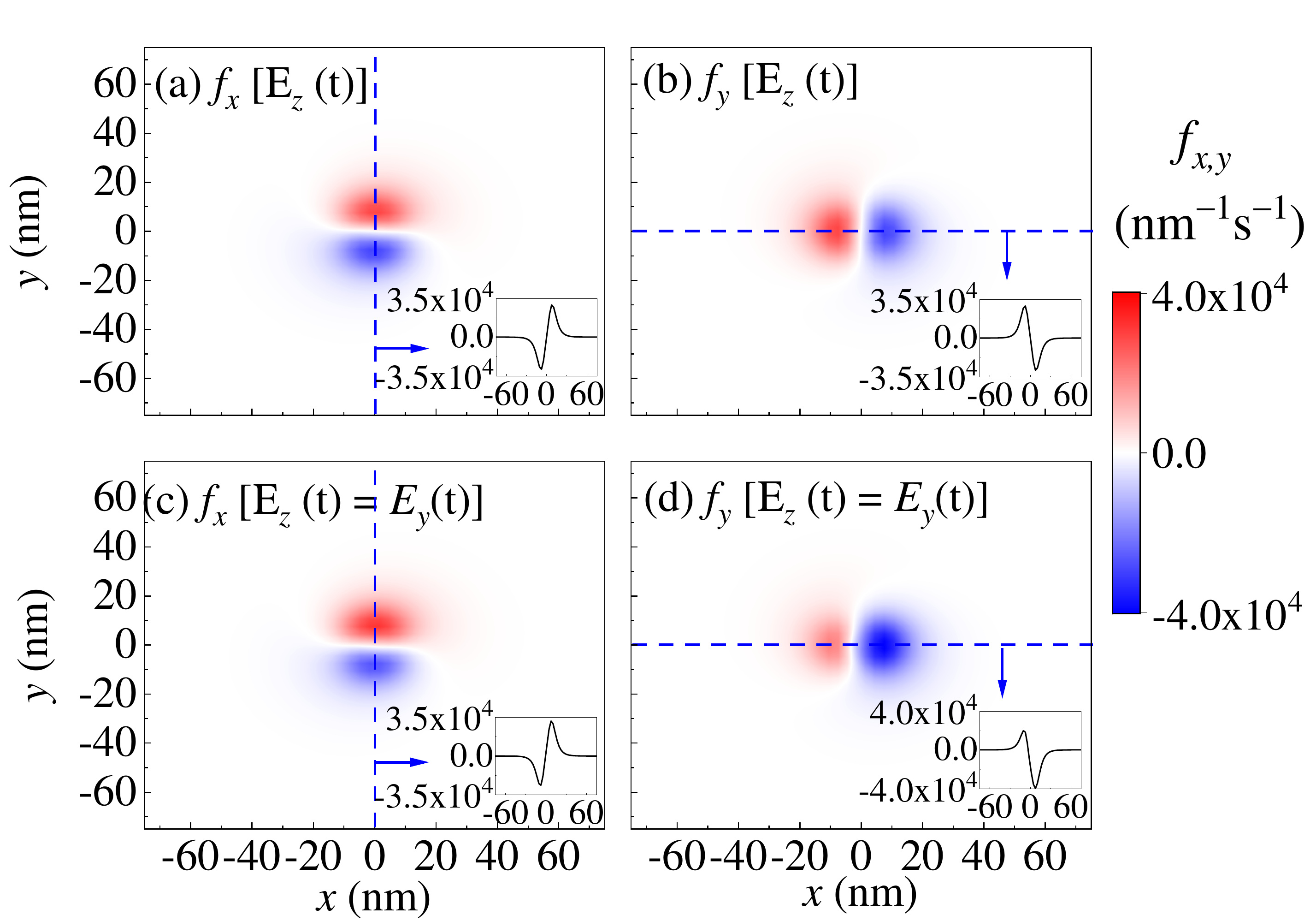}
	\caption{\label{torquef} Spatial profiles of the force density $ f_i = \gamma \vec{m}_s \cdot [\partial_i \vec{m}_s \times \langle \vec{m} \times \vec{H}_{\mathrm{\rm eff}} \rangle] $ for (a-b) $ E_z(t) $ without $ E_y(t) $ and (c-d) $ E_z(t) = E_y(t) $ (i.e. $ \phi = 0 $). Here, the microwave electric fields have the amplitude $ 1 $ MV/m and the frequency $ f = 2.825 $ GHz. The insets show the force density $ f_{x,y}$ profiles along the blue dashed lines. }
\end{figure}

Several peaks in the velocity vs. frequency curve can be found in the spectrum excited by $ E_z(t) $, and other peaks correspond to the $ E_y(t) $ component. To clarify the connection between the skyrmion velocity and skyrmion modes, we multiplied  the spectrum excited by $ E_z (t) $ and the spectrum excited by $ E_y (t) $,  see Fig. \ref{fre}(b), and achieved a better agreement between the spectrum and the $ |\vec{v}| ( f )$ curve. From this feature we infer  that combining two types of oscillation modes drives the skyrmion motion. For the analytic description of the skyrmion motion driven by the microwave electric field, we use a slow varying magnetization vector $ \vec{m}_s $ corresponding to the moving skyrmion profile and a fast vector $ \vec{n} $ for the GHz magnetization oscillation. Exploiting the ansatz $ \vec{m} = \vec{m}_s + \vec{n} $ for the  total magnetization vector in the LLG equation (\ref{LLG}), and following the procedure from Ref. \cite{PhysRevB.92.020403}, we deduce the Thiele's equation for the skyrmion motion:
 \begin{equation}
\displaystyle \vec{G} \times \vec{v} + \hat{D} \vec{v} = \vec{F},
\label{thiele}
\end{equation}
where we used the following notations: $ \vec{v} = (v_x, v_y) $, $ \vec{G} = 4 \pi Q \vec{e}_z $, and the tensor $ \hat{D}_{ij} = \alpha \int(\partial_i \vec{m}_s \cdot \partial_j \vec{m}_s )dx dy = \delta_{ij} 4 \pi \alpha $, where $ i,j = x,y $. We note that several terms such as $ \langle \vec{n} \times \dot{\vec{n}} \rangle $ and $ \langle \vec{n} \times \dot{\vec{m_s}} \rangle $ do not contribute to the Thiele's equation because their time average over the fast oscillations vanish. The driving force $ \vec{F} $ explicitly reads
 \begin{equation}
\displaystyle F_i = \gamma \int \vec{m}_s \cdot [\partial_i \vec{m}_s \times \langle \vec{m} \times \vec{H}_{\mathrm{\rm eff}} \rangle]dx dy.
\label{force}
\end{equation}
The skyrmion velocity is defined as follows:
\begin{equation}
\begin{aligned}
 v_{x1} = \frac{-\alpha F_x + F_y}{4 \pi Q}, v_{y1} = -\frac{F_x + \alpha F_y}{4 \pi Q}.
\label{velocity}
\end{aligned}
\end{equation}
Extracting the force density from the simulation results, we draw their spatial profiles in Fig. \ref{torquef}. If one applies $ E_z (t) $ (without $ E_y (t) $), the force density is symmetric, and the total force is zero. Therefore, the skyrmion does not move. Simultaneous application of the two components of the electric field  $ E_z (t) $ and $ E_y (t) $ leads to   asymmetric force distribution and a nonzero net force. Substituting the net force into Eq. (\ref{velocity}), we obtain the velocity $ (v_{x1}, v_{y1}) $ in a good agreement with the simulation results (Fig. \ref{phasevxy}(c)).

\section{Skyrmion motion driven by broadband THz pulse}

\begin{figure}
	\includegraphics[width=0.50\textwidth]{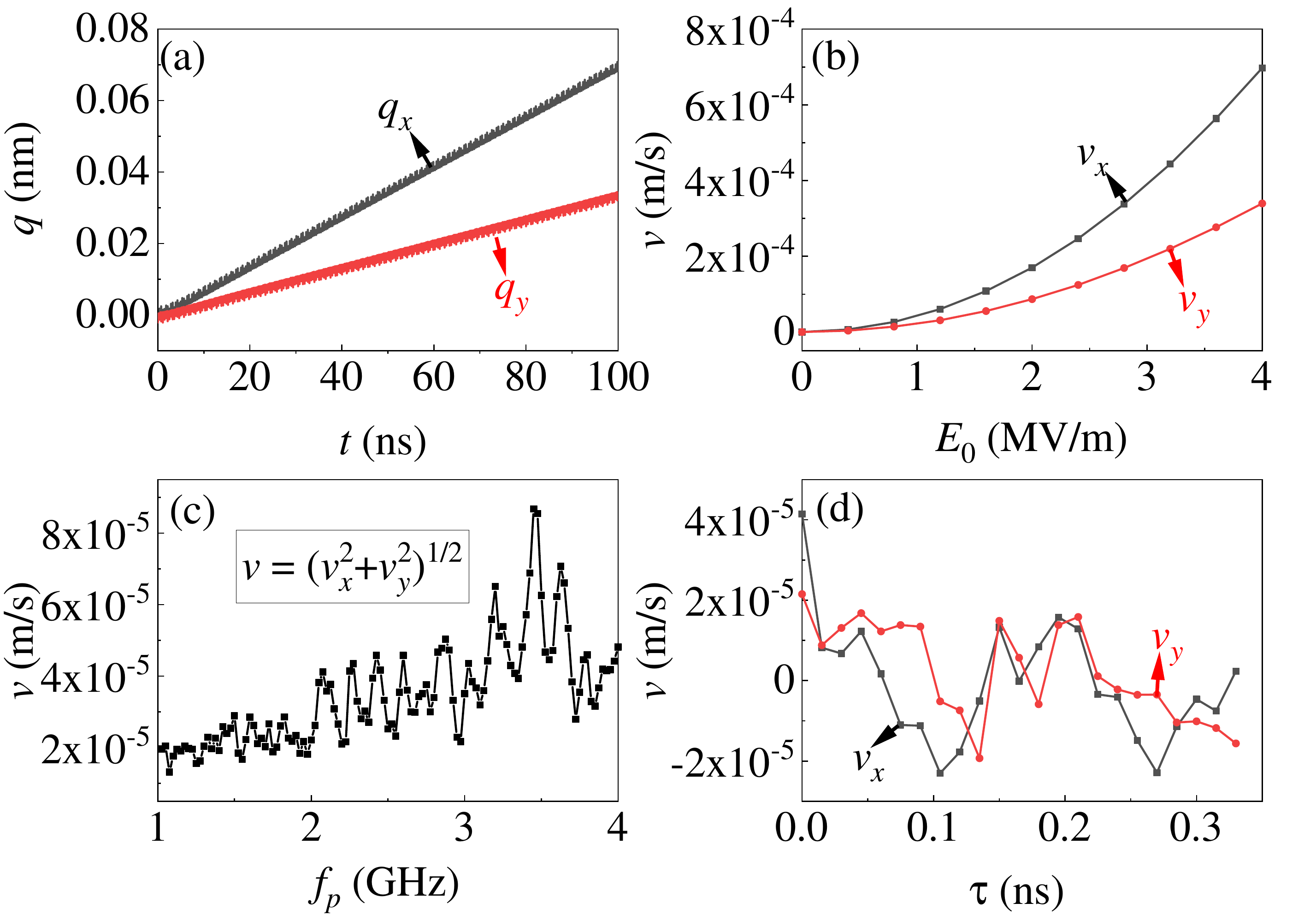}
	\caption{\label{pulse}  (a) The dynamics of skyrmion driven by two time-asymmetric pulses $ E_z(t) $ and $ E_y(t) $ with amplitude $E_0 = 4 $ MV/m, $ f_p = 1/T = 2.825 $ GHz and $ \tau = 0 $. (b-d) The velocity $ (v_x, v_y) $ as a function of the pulse amplitude $ E_0 $ (b), $ f_p = 1/T$ (c) and $ \tau $ (d). }
\end{figure}

For final results we apply a sequence of two   time-asymmetric broadband THz pulses with the period $ T $ and delay time $ \tau $ (Fig. \ref{model}(d-e)), where the length of the  pulse head is 15 ps (i.e., around THz). The employed pulses are experimentally feasible \cite{Hassan2016}. The asymmetric pulses have a strong head with a short duration, and a much longer and weak part in the opposite direction \cite{MOSKALENKO20171, PhysRevLett.94.166801}. Compared to the previous section and a single frequency excitation problem, the periodic time-asymmetric pulse carries many frequencies (integer multiple of the frequency $ f_p = 1/T $). Our simulations validated the fact that such a combination of two pulses $ E_z(t) $ and $ E_y(t) $ can govern the continuous motion of a skyrmion. As demonstrated in Fig. \ref{pulse}(a), for $ E_{z0}= E_{y0} = E_0 = 4 $ MV/m,  $ f_p = 1/T = 2.825 $ GHz and $ \tau = 0 $, the skyrmion velocities along $ x $ and $ y $ axis are positive. The velocity increases with the pulse amplitude $ E_0 $ quadratically (Fig. \ref{pulse}(b)). This result is in agreement with the result obtained by means of the microwave excitation (Fig. \ref{phasevxy}(d)). Changing the pulse period $ T $ one  changes the skyrmion speed, as shown in Fig. \ref{pulse}(c). Multiple peaks are observed in the $ v $ versus $ f_p $ curve, indicating  resonance features. Furthermore, the direction and speed of skyrmion motion are affected by the delay time $ \tau $ between $ E_y(t) $ and $ E_z(t) $, as demonstrated in Fig. \ref{pulse}(d). As distinct from  Fig. \ref{phasevxy}(c) for microwave excitation, in the case of pulses, the dependence of $ (v_x, v_y) $ on $ \tau $ is more sophisticated, as the skyrmion oscillations are excited by multiple frequencies.

\section{Conclusions}

Controlling skyrmion dynamics and skyrmion transport is a challenging problem in skyrmionics. This is especially important from the point of view of possible applications of skyrmions in information storage and information processing.
In the present work, we proposed the protocols based on skyrmion-driving with  GHz microwave harmonic and THz broadband pulses. We proved that an applied microwave electric field with both out-of-plane and in-plane components drags the skyrmion persistently.
In particular, the electric field's amplitude, frequency, and phase are critical parameters for controlling skyrmion motion. The effective torque that drives the skyrmion is traced back to  two different types of oscillations in the skyrmion magnetic texture. The results obtained through   micromagnetic simulations are supported by analytical considerations, and are directly relevant for skyrmionic-based  spintronic.

\textit{Acknowledgements}: The work is supported by Shota Rustaveli
National Science Foundation of Georgia (SRNSFG) (Grant
No. FR-19-4049), the National Natural Science Foundation of China
(Grants No. 12174452, No. 12074437 and No. 11704415), the Natural Science Foundation
of Hunan Province of China (Grants No. 2022JJ20050 and No. 2021JJ30784), the Central South University Innovation-Driven
Research Programme (Grant No. 2023CXQD036), and by the National
Science Center in Poland by the Norwegian Financial Mechanism
2014-2021 under the Polish-Norwegian Research Project NCN GRIEG
(2Dtronics) no. 2019/34/H/ST3/00515 (JB), and as a research Project No. DEC-2017/27/B/ST3/ 02881 (VKD),  and the DFG through SFB TRR227, and Project Nr. 465098690.

\bibliography{3nems}

\end{document}